\documentclass[12pt,preprint]{aastex}

\usepackage{tabularx}
\usepackage{url}
\usepackage{graphics,graphicx}
\usepackage{psfig}
\usepackage{tabu}
\usepackage{placeins}
\usepackage{rotating}
\usepackage{booktabs}

\begin{document}

\title{Using Random Forest Machine Learning Algorithms in Binary Supernovae Classification}
\author{Jonathan Markel\altaffilmark{1,2}, Amanda J. Bayless\altaffilmark{2,3}}
\altaffiltext{1}{The University of Texas at Austin, Department of Aerospace Engineering and Engineering Mechanics, 110 Inner Campus Drive, Austin, TX 78712, USA}
\altaffiltext{2}{Southwest Research Institute, Space Science and Engineering 6220 Culebra Rd. San Antonio, TX 78238, USA}

\altaffiltext{3}{The University of Texas at San Antonio, 1 UTSA Circle, San Antonio, TX 78249, USA}



\begin{abstract}
In the era of large all-sky surveys, there will be a need for rapid, automatic classifications of newly discovered transient objects.  Our focus here is the classification of supernovae (SNe). We consider random forest machine learning algorithm applied to the classification of Type Ia and core-collapse supernovae (CCSNe) by full light curves. We also quantitatively show the potential of early-epoch classification using the same machine learning techniques.  The algorithm uses the shape and magnitude of the light curve peak to determine the classification. This an initial study to essentially determine SN type with as few data points as possible.  New all-sky surveys will discover new transients at a rapid rate and decisions will have to be made on very little data which transients to follow-up. Here we present an initial study where as few as five data points near peak, we an identify whether it is a Type Ia or CCSNe with better than 80\% accuracy, but with several biases in the data that will require addressing in future work.

\end{abstract}



\section{Introduction} \label{sec:intro}
Supernovae (SNe) are the transient result of catastrophic star explosions. Astronomers discover several hundred SNe every year, and each one can provide valuable scientific insight into the progenitor systems and the environments in which they explode. The upcoming Vera C. Rubin Observatory's Legacy Survey of Space and Time (LSST) will observe the sky in six filters (\textit{ugrizy}) and is expected to detect 10 million SNe over its ten-year survey, or over 2000 newly discovered each night \footnote{\url{https://dmtn-102.lsst.io/DMTN-102.pdf}}. As SNe are visible for weeks to months, this will be a revolution in the number of SNe discovered and available for observation.  The most prolific SNe discovery surveys to date are the All-Sky Automated Survey for SuperNovae, which has discovered nearly 1000 bright SNe since 2014 \citep{holoien2017asas} and the Zwicky Transient Facility, which finds about 3-4 SNe nightly and more than 1000 in the past year \footnote{\url{https://www.ztf.caltech.edu/news/zwicky-transient-facility-nabs-several-supernovae-a-night}}. The massive amount of data from the current surveys and the upcoming LSST is a hurdle for astronomers who seek to prioritize follow-up observational resources. We note that the available data in early epochs will likely be photometric, and not spectroscopic.  This is a proof-of-concept study to distinguish between different types of SNe using only photometric observations, and to test early epoch performance of random forest machine learning algorithms for photometric classification. Shown in this paper, the initial results are promising and will improve with additional training sets and real, not synthetic, LSST data which are spectroscopically typed.  


\subsection{Supernovae Classification}
Despite a wide-ranging spectrum and light curve-based taxonomy, most SNe are believed to be either thermonuclear explosions of white dwarf stars (Type Ia SNe) or the collapse of a star's iron core (Type Ib/c and all Type II).  SNe are classified by their optical spectra into two types: Type II SNe exhibit hydrogen in their spectra and Type I SNe do not.  In general, Type Ia SNe show strong Si II spectra lines and occur at a rate of $0.301\times10^{-4}$ per year per Mpc \citep{li2011}. Interesting peculiar Type Ias include the Type Iax, which have weaker Si II lines and have an early hot, blue spectrum and are the most common peculiar Type Ia SNe \citep{jha2006}, and the 1991bg-like, subluminous Type Ias \citep{turatto1996}.  Type Iaxs make up the majority of peculiar Type Ias with approximately one-third of all Type Ias being Type Iaxs \citep{ryan2013}.  There is still  debate on the number of progenitor systems being single or double degenerate \citep{maoz2014, livio2018, jha2019}. 

In core-collapse SNe (CCSNe), Type Ib SNe exhibit He I lines while Type Ic SNe exhibit neither Si II or He I lines. Type Ib/c are likely derived from striped-envelope stars \citep{penko1997}. Furthermore, subtypes within Type II SNe may depend on either spectra and photometric light curves. Type IIn SNe are so named because they exhibit narrow Balmer emission spectrum lines \citep{schlegel1990} in early-epoch observations. Type IIb SNe display He signatures in later spectral observations. Type II-L and Type II-P are classified from their light curve characteristics \citep{penko2005}. After maximum brightness, Type II-L SNe decline “linearly,” and Type II-P SNe “plateau” for an extended period (a complete review of SNe subtypes can be found in \citet{penko1997}). CCSNe occur at a rate of $0.72\times10^{-4}$ per year per Mpc with Type IIPs comprising 52.4\% of CCSNe, Type IIL being 7.3\%, Type IIns at 6.4\%, Type Ib comprising 6.9\%, Type Ic being 17.6\%, and the rest assorted faint and peculiar SNe \citep{strolger2015}.

We need to observe SNe at as early of an epoch as possible for several reasons. By observing at early epochs, we can better understand the evolution of the shock interactions. The shock dynamics shows valuable information about the physical properties of the progenitor stellar environment and in massive stars any mass loss material the star ejected into the surrounding adding to the interstellar medium \citep{ro2013shock}.  The shock interactions in all types of SNe occur at early times and it is  difficult to reconstruct in models with only late time observations. Also, any follow-up observations should include space-based UV observatories.  In CCSNe the shock interactions are quite bright in the UV at early times \citep{sapir2017uv}. \citet{pritchard2014bolometric} also demonstrated that the bolometric luminosity in CCSNe is underestimated by more than 50\% if the early-time UV observations are not included. 

Secondly, in Type Ias early-time Kepler data shows a more complicated rise to peak in the first five days (\citealt{dimitriadis2019}, SN 2018oh). The early fast rise could be the interaction with the companion star.  A similar blue bump was seen in SN 2017cbv and interpreted as an interaction with a sub-giant companion \citep{hoss2017}.  However, these interpretations are model dependant.  With more observations of Type Ias within the first week, we could distinguish between companion interactions or other models, such as a helium shell or an uneven Ni distribution \citep{dimitriadis2019}.  Additional information can be obtained from the first 5-day color evolution. \citet{bulla2020} show a correlation between the {\it g-r} slope and the brightness of the Type Ia SNe, but all coming from one population.  Conversely, \citet{stritzinger2018} see two populations in the early (B-V) evolution, a fast rising redder population and a slow rising blue population. Cosmolagically, there is an interest in calibrating Type Ia luminosity to better refine these as a "standard candle"  For example, the \citet{hayden2019} model correlates the luminosity with the rise and fall of the light curve. However, \citet{hayden2019} note that the fast rise time to peak will require early detection and early observations at a faster observing cadence than has been done in the past.  It is clear that early detection and observation is going to be crucial moving forward to determine SNe mechanisms of all SNe types. 

\subsection{Machine Learning Process and Motivation}\label{sec:lit_review}
A literature review revealed three key findings that guided our study:

\begin{enumerate}
    \item Machine learning algorithms are very well suited to analyzing the large amounts of astronomical data expected from the large surveys and missions \citep{borne2008machine, karpNN2013, kessler2010results, lochner2016, moller2016, newling2011, richards2011semi}.
    
    \item Large data sets of synthetically generated SNe light curves or spectra are necessary to train machine learning classifiers \citep{kessler2010supernova, lochner2016, moller2016, newling2011}.
    
    \item The majority of research into SNe classification considers photometric observations from the beginning to the end of the explosion, and very few classify SNe during the first several observations \citep{kessler2010results}.
    
\end{enumerate}

Finding 2 guided the selection of data used to train the classification algorithm. Synthetic data is necessary as training requires 100s to 1000s of well sampled data sets. While 1000s of SNe have been discovered, there is not enough follow-up data for each type to use observations as the training set. Additionally, SNe phototometric and spectral observations are obtained with a``best-effort" approach to observe as many as possible, often without any \textit{a priori} information. Consequently, many rare types of SNe (e.g. Type IIns) are missed and there is a bias towards observing bright SNe. Thus, there is not necessarily a representative sample of SNe types in spectroscopically confirmed observations, which would result in a biased classifier.  

Finding 3 was especially significant to us. The earlier a rare SNe can be identified, the greater the chance that it can be measured with spectroscopy leading up to and during its peak magnitude. Spectroscopy during peak versus other epochs is useful as the SN is simply brighter making spectral integration times shorter and have a higher signal to noise.  This would be an optimal time to spectroscopically type most SNe and estimate the ejecta velocity.  By observing the SN time-series spectrum, we are able to observe the evolution of features, such as the temperature (assuming thermal blackbody emission).  SNe are very UV bright in the first 5 days, but then fade quickly from \ion{Fe}{2} line blanketing (e.g. SN 2012aw, \citealt{bayless2013}). Additionally, time-series spectroscopy that included early-time spectra allowed the discovery of Type IIb SNe as a distinct type from the Type Ib SNe (e.g. SN 1993J, \citealt{nomoto1993}).  The motivation is to identify SNe types as early as possible, using only light curves (as spectra will likely be unavailable), with as few data points and as early as possible.  Having a first guess as to the type, we can then inform the large and space-based observatories which of the many SNe to observe for follow-up observations. 

\section{Methodology}\label{sec:method}
In this section we discuss the classification algorithm used, including inputs (features), performance metrics, and implementation. We apply the same algorithm data in two different cases: 1) classification using all the available light curve data and 2) using only the first few measurements (early-epoch classification). Additionally, we discuss the photometric light curve data used to classify SNe, including methods for parsing and potential biases of simulated SNe data.

\subsection{Data}\label{sec:data}

Several existing transient classification research papers utilized the 2010 Supernovae Classification Challenge (SNPhotCC) data set for testing \citep{karpNN2013, newling2011, revsbech2018}. There is also a new synthetic dataset generated specifically for LSST, the Photometric LSST Astronomical Time-Series Classification Challenge (PLAsTiCC; \citealt{plasticcdata, kessler2019models}).  Both the SNPhotCC and PLAsTiCC use the SDSS and CSP data to construct templates. It is noted in section 4.4.2 of \citet{kessler2019models} that the Type II models used for SNPhotCC had artifacts in the {\it z} and Y bandpasses, which were corrected in the PLAsTiCC dataset.  This would likely skew the identification of Type II and cause a higher importance in the redder bandpasses. Therefore, we must be careful using the SNPhotCC data as it will likely give an optimistic result. Nevertheless, we utilized the SNPhotCC for several reasons.  Preparatory materials for PLAsTiCC recommend utilizing the SNPhotCC dataset for building the software framework, which was started in this study prior to the availability of the PLAsTiCC data.  Secondly, we wanted to compare our results to that of \citet{lochner2016, moller2016}.  This comparison is discussed in Section 4 and the algorithms do, in fact, prefer the redder band passes, which was also the case in \citep{lochner2016}.   

The SNPhotCC data consists of 21,318 simulated light curves and is the most extensive and available representative datasets of SNe, making it especially useful for testing machine learning classification algorithms. It contains simulated Type Ia, Ib/c, and II SNe observations in the \textit{griz} filters based on the Dark Energy Survey (DES). Non-type Ia SNe simulations are based on spectrographically confirmed light curves donated by the Carnegie Supernova Project, the Supernova Legacy Survey and the Sloan Digital Sky Survey.  The contribution of non-Ia data from these surveys likely undersamples the variety of SNe that will be observed by LSST, but “will be a useful step away from the overly-simplistic studies that have relied on a handful of non-Ia templates” \citep{kessler2010supernova}. 

We also had access to the Supernova Analysis Application (SNAP) database \citep{bayless2017supernovae}, which currently contains several hundred observed SNe from space and ground observations\footnote{\url{https://snap.space.swri.edu/snap/}}. However, the SNPhotCC dataset has several advantages over SNAP for this area of research. First, the SNAP database contains SNe from multiple surveys, each with slightly different sets of filters, and this non-homogeneity presents special challenges when comparing the same SNe metrics across different sky surveys. More importantly, the literature review and large number of observations expected from LSST led us to conclude that machine learning algorithms would be best suited for the scale of autonomous SNe classification task, and machine learning algorithms require large amounts of data to adequately train. While the number of SNe in SNAP is notable, we opted to use a dataset of synthetically generated SNe to train the classifier on a greater variety of types.

\subsubsection{Data Cleaning} \label{sec:cleaning}
Raw data of astronomical observations often come with noise and other factors that must be accounted for. Additionally, an astronomical transient may not always be observed in every filter. 
We utilize random forests in our research. However, the scikit-learn implementation of random forests does not handle missing values, thus missing data must be imputed without biasing the model.  After calculating values for each SNe feature, we impute missing values in the training set with the median value for each column of non-zero observations per \citet{rfclassificationwebsite}.

We acknowledge that imputation with values dependent on the training set population distribution (e.g. median, mean) can be unstable when the training set is non-representative. However, median imputation was utilized here to achieve initial results, with the plan for further development to address this in more detail. Future work should take care to test a chosen imputation scheme with rigorous k-fold cross-validation.

\subsection{Random Forest Classification}\label{sec:random}
Boosted decision trees (e.g. XGBoost; \citealt{chen2016xgboost}) and random forest classifiers have been shown to be one of the most robust algorithms for photometric classification when compared to naive Bayes, \textit{k}-nearest neighbors, support vector machines, and convolutional neural networks \citep{lochner2016}. Decision trees map input features to output classes based on specific rules. The model is then trained by recursively selecting features and boundaries which increase the overall information gain. In other words, decision trees find boundaries which minimize the entropy of the working set. A single decision tree has its drawbacks –-- it can be prone to over-fitting and sensitive to small changes in the input data. However, these effects can be mitigated by combining multiple decision trees into an ensemble, where the final classification is determined by a “majority vote” of the many trees contained within the model. A key benefit of ensemble decision tree algorithms like random forests is that the confidence of any classification can be directly interpreted from the proportions of decisions of individual trees. We utilized the scikit-learn random forest classifier \footnote{\raggedright\url{https://scikit-learn.org/stable/modules/generated/sklearn.ensemble.RandomForestClassifier.html}}, an ensemble method that combines many decision trees. 

\subsubsection{Model Parameters}
The number or estimators, or trees, in a random forest classifier can impact the models performance. Random forests cannot be overfit \citep{breiman2001random}, but after a certain number of trees, increasing the number of estimators has a detrimental effect on computational runtime and does not improve performance. However, more estimators do increase the stability of the model. Individual trees in a random forest are constructed using subsets of the total data. For each tree, testing performance on the unused subset produces an internal error estimate. Combining these internal error estimates creates the out-of-bag error estimate \citep{breiman1996out}, which has comparable accuracy to using a test set the same size as the training set \citep{breiman1996bagging}. 

\begin{figure}[ht]
    \centering
    \includegraphics[scale=0.4]{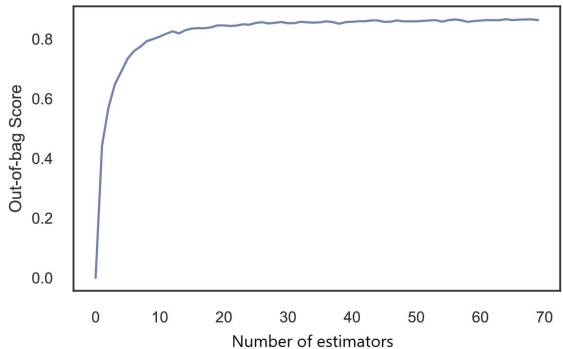}
    \caption{Comparison of out-of-bag classifier performance for varying numbers of trees.}
    \label{fig:estimators}
\end{figure}

We determined the optimal number of trees for our classifier by calculating the out-of-bag score (equal to 1 --– [out-of-bag error]) for models ranging from 1 to 70 estimators. We implement the out-of-bag error using the ``oob score" attribute of the scikit-learn RandomForestClassifier \citep{scikit-learn,hastie2009}. After analyzing performance trade-offs in Figure \ref{fig:estimators}, we selected 40 trees. 

\subsection{Feature Selection}\label{sec:features}
We consider two techniques for feature extraction. 
In the first method, we considered a general parametrization fitted with the following form, which we refer to as the ``Bazin Form'':

\begin{equation}
f^k(t) = A^k \frac{e^{-(t-t_0^k)/ \tau_{fall}^k}}{1 + e^{-(t-t_0^k)/\tau^k_{rise}}} + c^k
\end{equation}

where t$_{fall}$ and t$_{rise}$ are the times before and after peak magnitude, \textit{c} is a constant, A normalizes the signal, and t$_{0}$ relates the rise and fall times to the time of peak magnitude \citep{bazin2009, dai2018lsstsupernovae}. The five Bazin fit parameters do not represent any physical characteristics but are sufficiently general to fit all SNe types. This model is used in other photometric classification literature to reduce the dimensionality of light curve data \citep{karpNN2013,moller2016,dai2017cosmology, ishida2019}.  

\begin{figure}[ht]
    \centering
    \includegraphics[scale=0.55]{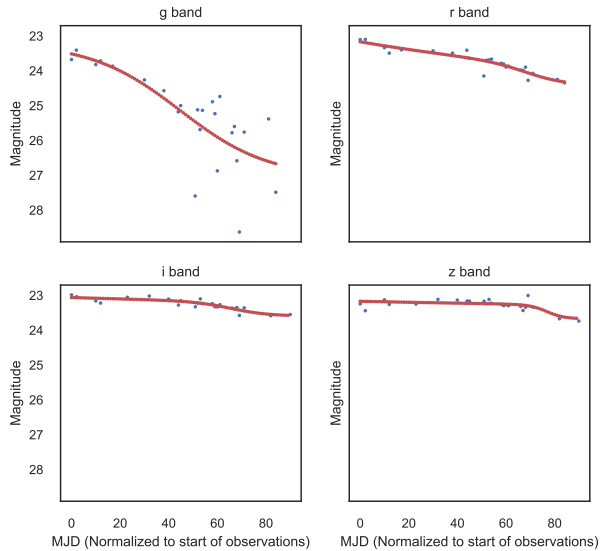}
    \caption{An example of an acceptable fit using the Bazin model.}
    \label{fig:fit_good}
\end{figure}

\begin{figure}[ht]
    \centering
    \includegraphics[scale=0.55]{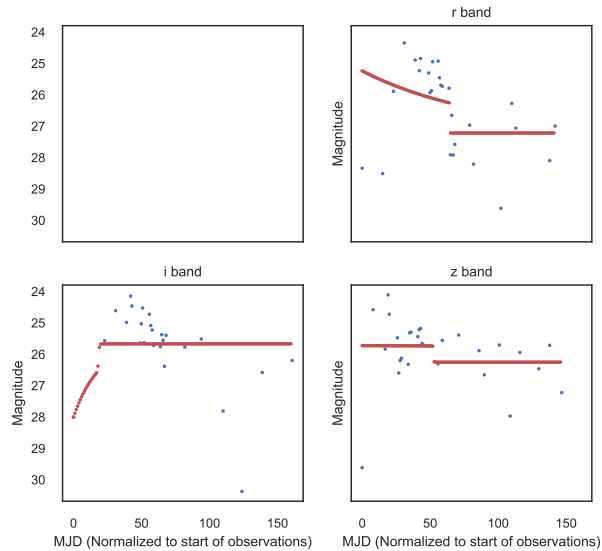}
    \caption{An example of an unacceptable fit using the Bazin model.  The {\it g}--band is missing in this case.}
    \label{fig:fit_bad}
\end{figure}

In Figures \ref{fig:fit_good} and \ref{fig:fit_bad}, we show examples of both acceptable and poorly fitted data.  In Figure \ref{fig:fit_good}, the model does decently representing the underlying data points across the \textit{g}, \textit{r}, and \textit{i} bands. However, the fit does not perfectly converge in the \textit{z} band. 

In Figure \ref{fig:fit_bad}, none of the fits adequately represent the underlying data. Additionally, this SN lacks observations in the \textit{g}--band. This does not have an effect on fitting the other bands, as the underlying model fits each filter independently.  In this first modeling attempt, the Bazin model tends to favor a low varience in the light curve. However, it still produced an acceptable fit in the {\it g}--band for this SN even with late-time scatter.  

In a second method, we describe light curves using statistics that measure the variability and characteristics of a light curve. This technique consists of using six features to describe the curve in each filter, namely maximum magnitude, kurtosis, skew, median absolute deviation, the Shapiro-Wilk statistic \citep{shapiro1965}, and the coefficient of variation (Table \ref{table:features}). 
Each feature was normalized by removing the mean and scaling to unit variance. We use these parameters to implicitly reflect light curve characteristics which are known to be indicators of physical parameters of core-collapse SNe (CCSNe), such as the explosion energy, envelope composition, or mass.  

\begin{table}[ht]
\centering
    \begin{tabular}{ | m{1.5cm} | >{\centering\arraybackslash}m{2.7cm}| m{2.5cm}| } 
    
        \hline
        Value & Formula & Feature Description\\
        \hline
        Maximum Peak & $ max(A) $ & Peak magnitude value \\ 
        \hline
        Kurtosis & $ n \frac{ \Sigma_{i=1}^n (X_i - X_{avg})^4 } { (\Sigma_{i=1}^n (X_i - X_{avg})^2)^2 } $ & "Sharpness" of distribution \\
        \hline
        Skew & $\frac{\bar{X} - Mo}{\sigma} $ & Distribution asymmetry measure \\
        \hline
        Medium Absolute Deviation & $ median(X_i - median(X))$ & Measure of population variability \\
        \hline
        Coefficient of Variation (CV) & $ \frac{\sigma}{\bar{X}} $ & Normalized variability to mean relation \\
        \hline
        Shapiro-Wilk Statistic & $ \frac{(\Sigma_{i=1}^n a_i x_{(i)})^2}{\Sigma_{i=1}^n(x_i-\bar{x})^2} $ & Measure of distribution normality \\
        \hline
    \end{tabular}
    \caption{Full light curve statistical features used in classification.}
    \label{table:features}
\end{table}

\subsection{Early-Epoch Classification}
The SNPhotCC gave participants the extra challenge to classify SNe using only the first six observations in any filter, however, it was not attempted by any participants due to time constraints and increased interest in the full light curve challenge \citep{kessler2010results}. Given our interest in early-epoch classification, we use the challenge constraint of six points as a baseline. We add an additional constraint for our early-epoch analysis -- being able to determine type before the SNe reaches it's greatest apparent magnitude. However, some simulated SNe in the SNPhotCC have fewer than 6 observations before the peak magnitude due to the edge effects of simulating Dark Energy Survey data. The SNPhotCC also considers the time constraints of SN surveying, and not all light curves start at explosion.

In our early-epoch analysis of the the SNPhotCC data, we identify the peak magnitude in each light curve and applied a quality cut wherein any SNe with fewer than five data points before the peak in two or more filter bands are removed. This rule results in the removal of 5,609 SNe, or 26.3$\%$ of the data set. For the approximately 15,700 remaining SNe, all post-peak observational data was discarded. Each filter band was then fit to a quadratic polynomial, and filter bands with fewer than three points were fit to a line. 

Many of the statistical features used to analyze full light curves are dependent on a full distribution with an identifiable peak. With our early epoch data set, we cannot assume knowledge of a defined peak magnitude. A reduced set of statistical features and one new feature were used, and can be found in Table \ref{table:earlyepochfeatures}. The early-epoch classification portion of this research includes two cases -- one in which the statistical features are calculated from discrete observational data, and another where the data is fitted to a third-order polynomial. 

\begin{table}[ht]
\centering
    \begin{tabular}{ | m{1.4cm} | >{\centering\arraybackslash}m{2.7cm}| m{2.2cm}| }
        \hline
        Value & Formula & Feature Description\\
        \hline
        Average Rate of Change & $ \frac{\Sigma_{i=1}^n (x_{i+1} - x_{i})}{n} $ & Slope of the rise to peak\\
        \hline
        Coefficient of Variation (CV) & $ \frac{\sigma}{\bar{X}} $ & Normalized variability to mean relation\\
        \hline
        Skew & $\frac{\bar{X} - Mo}{\sigma} $ & Distribution asymmetry measure \\
        \hline
        Kurtosis & $ n \frac{ \Sigma_{i=1}^n (X_i - X_{avg})^4 } { (\Sigma_{i=1}^n (X_i - X_{avg})^2)^2 } $ & "Sharpness" of the distribution \\
        \hline
    \end{tabular}
    \caption{Early epoch classification features.}
    \label{table:earlyepochfeatures}
\end{table}

\subsection{Performance Metrics}\label{section:performance metrics}
To measure the performance of our model, we use standard performance metrics such as precision, recall, and the F1 score \citep{f1rijsbergen}, which is the harmonic mean of precision and recall. Precision ($ Precision = \frac{TP}{TP+FP}$) is the number of true positives (TP) divided by the combined true positives and false positives (FP), and it can be thought of as a measure of the classifier’s exactness. The recall ($Recall = \frac{TP}{TP+FN}$) is the number of true positives divided by the combined true positives and false negatives (FN). Recall can be thought of as the classifiers ability to classify the full set of positives. Additionally, positive class recall is referred to as the “sensitivity,” and negative class recall is referred to as “specificity.” The F$_{1}$ score is the weighted average of recall and precision \citep{yang}, and is calculated as $F_{1}(r,p) = \frac{2rp}{r*p}$. The F$_{1}$ score is more useful than standard metrics when the classes are unevenly distributed, as they are in the SNPhotCC data.

\section{Results}

One advantage of random forests is the ease with which the relative importance of any given feature for correctly classifying a SNe can be determined. The implementation of random forests in scikit-learn includes a function for determining a features importance. However, \citet{strobl2007bias} shows that the importance of features can be biased. Instead of using the built in feature importance function, we determine drop column importance for each feature \citep{parr2018beware}. The drop column importance of a feature is calculated by measuring the decrease in classification performance when a given feature is removed from the model. If the model performance decreases significantly without the feature, it is assigned a higher importance. It should be noted that collinear features can have an affect the interpretability of feature importance metrics, and future research should be careful to employ a feature correlation study and dimensionality reduction step to mitigate these effects. Additionally, we introduce a column of randomly generated numbers to help determine whether any given feature is contributing to classification performance. By comparing feature importance to a set of randomly generated numbers as a baseline, we can identify useless and even detrimental features to be those which are less important the random number set. To visualize the importance of features we normalize their importance relative to the importance of the randomly generated number set. The feature importances are normalized according to Equation \ref{eq:normalization}. $X$ is the drop column importance value calculated with the code supplied in \citet{parr2018beware}, $X_{random}$ is the importance of the set of randomly generated values, and $\sigma_{X}$ is the standard deviation of the set of feature importances.

\begin{equation}\label{eq:normalization}
    X' = \frac{X - X_{random}}{\sigma_X}
\end{equation}

We compare the performance of random forest classifiers when applied to full light curves and pre-peak light curves. Additionally, we classify SNe without interpolating data over a fitted model to provide a baseline performance. In all, we consider four variations:

\begin{enumerate}
    \item Full light curves, where 6 features are calculated from discrete data points in each filter.
    \item Full light curves, where 6 features are calculated from interpolated light curve data.
    \item Pre-peak light curves, where 4 features are calculated from discrete data points in each filter.
    \item Pre-peak light curves, where 4 features are calculated from interpolated light curve data.
\end{enumerate}

In both early-epoch and full light curve cases, in can be seen that interpolation of data -- even when not perfectly fit to a model -- improves classification performance. 

\subsection{Full Light Curve Classification}

Comparing classification performance between raw data (Table \ref{table:metrics_unfit_full}) and fitted data (Table \ref{table:metrics_fit_full}), it can be seen that interpolating data is a highly-effective way of increasing classifier capability. Specifically, we see an improvement in core-collapse SNe recall, and Type Ia precision. In other words, with fitted data, our classifier correctly typed a greater portion of core-collapse SNe and was more capable of correctly classifying Type Ia supernovae. 

\begin{table}[h!]
\centering
    \begin{tabular}{ |c|c|c|c| } 
         \hline
          & Precision & Recall & F$_{1}$ Score \\ 
         \hline
         Type Ia & 0.71 & 0.49 & 0.58\\ 
         Core Collapse & 0.85 & 0.93 & 0.89\\ 
         \hline
         Average/Total & 0.81 & 0.82 & 0.81\\
         \hline
    \end{tabular}
    \caption{Classification performance for \textit{unfitted}, full light curves.}
    \label{table:metrics_unfit_full}
\end{table}

\begin{table}[h!]
\centering
    \begin{tabular}{ |c|c|c|c| } 
         \hline
          & Precision & Recall & F$_{1}$ Score \\ 
         \hline
         Type Ia & 0.77 & 0.49 & 0.60\\ 
         Core Collapse & 0.85 & 0.95 & 0.90\\ 
         \hline
         Average/Total & 0.83 & 0.84 & 0.82\\ 
         \hline
    \end{tabular}
    \caption{Classification performance for \textit{fitted}, full light curves.}
    \label{table:metrics_fit_full}
\end{table}

\begin{figure}[ht]
    \centering
    \includegraphics[scale=0.6]{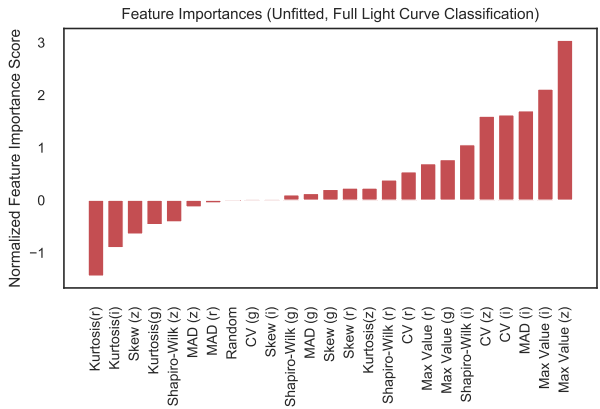}
    \caption{Drop column importance of features in a classifier trained on unfitted data.}
    \label{fig:importance_unfit}
\end{figure}

\begin{figure}[ht]
    \centering
    \includegraphics[scale=0.6]{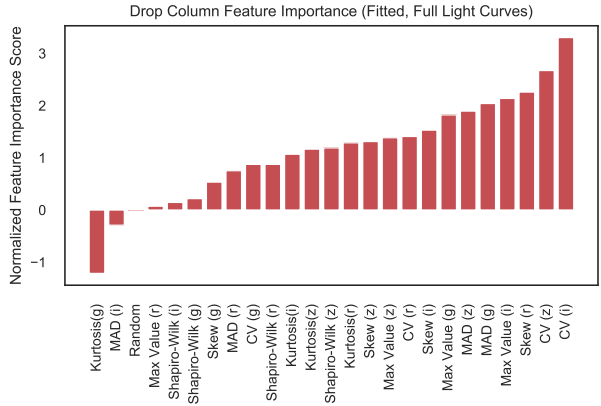}
    \caption{Drop column importance of features in a classifier trained on fitted data. Note decreased randomness of features when compared to a model trained on unfitted data.}
    \label{fig:importance_fit_full}
\end{figure}

We normalized the importance of individual features relative to a ``feature'' of randomly generated numbers. By fitting our data to a model and interpolating, we greatly reduce the overall randomness of features used for classification (compare Figures \ref{fig:importance_unfit},and \ref{fig:importance_fit_full}). Certain features are particularly important for correctly classifying a SN from light curve data, like the peak magnitude value. While the magnitude in the \textit{z} and \textit{i} filters were more important for fitting unfitted data, the \textit{g} and \textit{i} filters were more important for classifying SN with fitted light curves. From the importance of the peak magnitude and variation across filters, we infer that the peak magnitude is key to classifying SN no matter which observing filter is used.

Furthermore, when classifying actual SNe, one should remove any feature which is less important than the random feature. They are irrelevant at best and potentially harmful to the result.

\subsection{Early-Epoch Classification}

We see the same improvement in performance for early-epoch classification after fitting our data to a third-order polynomial. Comparing Tables \ref{table:metrics_unfit_ee} and \ref{table:metrics_fit_ee} shows an increase in core-collapse precision, while recall remains steady. This is in contrast to the full light curve classification task.

\begin{table}[ht]
\centering
    \begin{tabular}{ |c|c|c|c| } 
         \hline
          & Precision & Recall & F$_{1}$ Score \\ 
         \hline
         Type Ia & 0.61 & 0.28 & 0.38\\ 
         Core Collapse & 0.79 & 0.94 & 0.86\\ 
         \hline
         Average/Total & 0.74 & 0.77 & 0.74\\ 
         \hline
    \end{tabular}
    \caption{Classification performance for \textit{unfitted} early-epoch light curves.}
    \label{table:metrics_unfit_ee}
\end{table}

\begin{table}[ht]
\centering
    \begin{tabular}{ |c|c|c|c| } 
         \hline
          & Precision & Recall & F$_{1}$ Score \\ 
         \hline
         Type Ia & 0.72 & 0.45 & 0.55\\ 
         Core Collapse & 0.84 & 0.94 & 0.89\\ 
         \hline
         Average/Total & 0.81 & 0.82 & 0.80\\
         \hline
    \end{tabular}
    \caption{Classification performance for \textit{fitted} early-epoch light curves.}
    \label{table:metrics_fit_ee}
\end{table}

\begin{figure}[ht]
    \centering
    \includegraphics[scale=0.6]{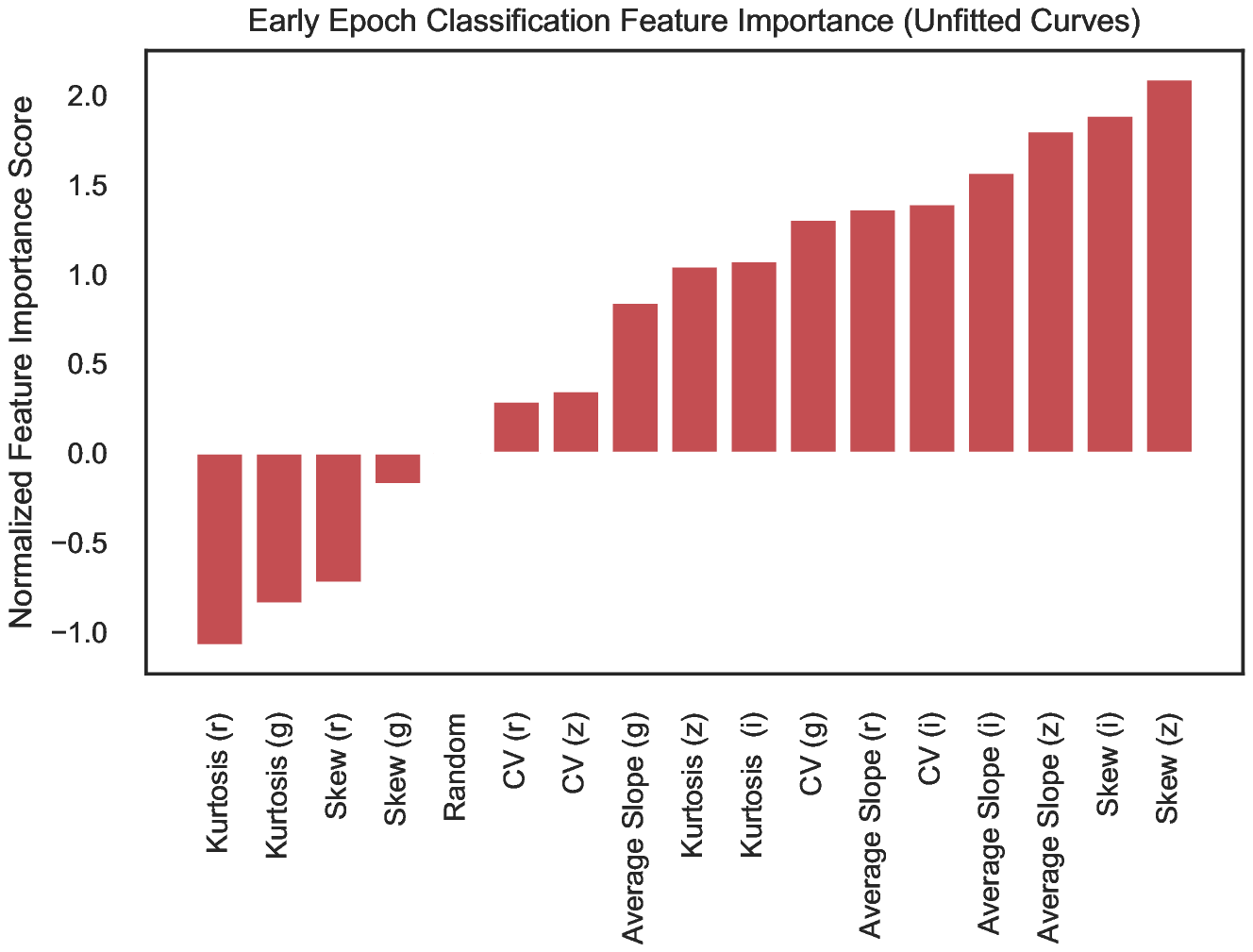}
    \caption{Drop column feature importance of unfitted early-epoch classification.}
    \label{fig:importance_unfit_ee}
\end{figure}

\begin{figure}[ht]
    \centering
    \includegraphics[scale=0.6]{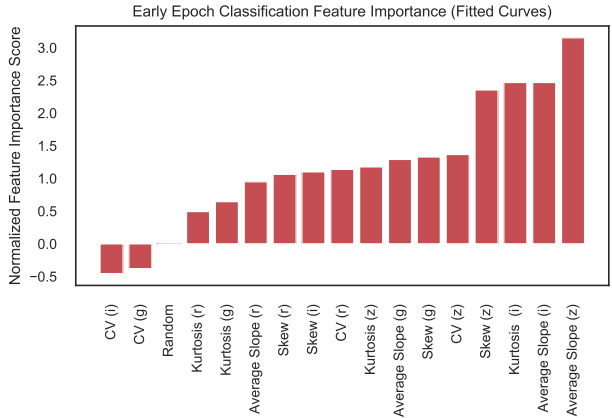}
    \caption{Drop column feature importance of fitted early-epoch classification.}
    \label{fig:importance_fit_ee}
\end{figure}

Figures \ref{fig:importance_unfit_ee} and \ref{fig:importance_fit_ee} confirm that the rate of rise to peak can be a useful indicator for early-epoch SN classification. In the absence of a defined peak, the skew of the data was highly important for both the unfitted and fitted data. We also see a decrease in feature randomness after fitting and interpolating data, similar to the full light curve case.

\section{Discussion}
In this paper we discuss our process of a creating a machine learning classifier for SNe. We successfully identified key areas for continued research, potential pitfalls for researchers getting started in astrophysical classification, and established a software framework for our future work to build upon. This is the start of a more targeted approach to identify SN types, particularly rare types, with limited information. This is difficult but necessary as there will be few light curve data points observed by LSST in early epochs.

\subsection{Necessity of Interpolating Light Curves}
The greatest challenge we faced in our research was fitting unclassified SNe models to an underlying model. 

The model proposed by \citet{bazin2009} is general enough to fit most SNe, which is especially beneficial when the target SNe’s type is unknown. However, a significant number of light curves fit with this model either did not converge or were not fit well enough to describe the underlying observations (see Figure \ref{fig:fit_bad}). As a result, SNe which did not fit the Bazin model had to be removed or fit with another model. Poor fits may be partially mitigated by iteratively refitting with different parameters, as applied in \citet{dai2018lsstsupernovae}. One area of future research is to combine the iterative approach applied by \citet{dai2018lsstsupernovae} with type-specific models. For example, once a SNe is fit to the general model and an initial guess at type is found, the SNe could be fit to a number of type-specific models, depending on classification uncertainty, in an attempt to improve classification. Once the best fit is found, the data could be interpolated, reclassified, and used to improve future predictions. This approach would also have the advantage of provide an automated, statistical means of validating type-specific models with each spectroscopically confirmed SNe.

\subsection{Feature Importance}
We compare our study here with that of other similar studies mentioned prior.  \citep{lochner2016}  test five different algorithms: naïve Bayes, k-nearest neighbors, multi-layer perception, a support vector machine, and boosted decision trees (BDT). For \citep{lochner2016} the BDT was the most effective with the important features focus on color and shape of the light curve, with redshift estimation being important if unknown from other means.  The time of the fall off seems to be their most important shape feature, and also in the redder band passes. They have an area-under-curve (AUC) between 80-90\% for the BDT calculation for a 90\% purity and 84\% completeness.
\citep{moller2016} also used a BDT, specifically XGBoost being the best, for their algorithm, with primarily a focus on Type Ia classification.  \citep{moller2016} found that the accuracy of their classifying was redshift dependent.  However, they have an efficiency of $\approx$40\% at a $\approx$95\% purity in identifying the Type Ias at all redshift.  This efficiency improved to 65-70\% at lower redshifts (0.2-0.4).  We are not getting as good of fits as these other groups, likely due to our choice of pre-processing and feature extraction.  However, we can make a few general conclusions. 

From the cases studied, we find that not all features are created equal in random forest classification. The peak magnitude is a key to classifying full light curves, as the peak magnitude value was a highly important feature (see Figures \ref{fig:importance_unfit} and \ref{fig:importance_fit_full}). Future researchers should consider the value in selecting unique features for each filter band, dependent upon the data being analyzed and the performance of their own model. It is highly recommended to remove any feature which is outperformed by a column of randomly generated numbers. 

Furthermore, our study of feature importance underscores the need for fitting data to a model. For example, in full light curve classification, comparing Figures \ref{fig:importance_unfit} and \ref{fig:importance_fit_full} shows how even inconsistent fitting drastically reduces the randomness in a model. The same is true for the early-epoch case, when most SNe were fit to a line or spline.

\section{Conclusion}
As computational power and the scale of astronomical surveys continue to improve, machine learning algorithms -- and especially random forest classifiers -- can be a robust and useful tool for understanding our universe. When used for transient classification, machine learning can help to quickly identify events worth closer examination. As our ability to identify SNe improves through the use of new technologies and methods, so will our understanding of the physical mechanisms behind these fundamental astronomical events.
\acknowledgments
  This work is supported by the NASA Astrophysical Data Analysis Program (NNH15ZDA001N-ADAP).  We would like to thank the anonymous referee for their comments. This project was limited to the scope of a summer undergraduate internship and we unfortunately could not address all of the referee's comments for publication in the scope allowed. We have edited this version to address some of the referee's comments and have made statements to acknowledge a first attempt at this coding where we could not. 

\bibliographystyle{plainnat}
\bibliography{biblio}

\end{document}